\documentclass[12pt]{article}
\usepackage{amsmath,amssymb,graphicx} 

\usepackage{psfrag}
\usepackage{subfigure}
\usepackage{tabularx}
\usepackage{footmisc}
\parskip=\baselineskip

\newcommand{\ZZ}{\mathbb{Z}}

\newcommand{\myfig}[3]{
	\begin{figure}[ht]
	\centering
	\includegraphics[width=#2cm]{#1}\caption{#3}\label{fig:#1}
	\end{figure}
	}

\newcommand\cc[1]{#1^{^{\kern-6pt \circ}}\kern2pt}

\def\pa{\partial}
\renewcommand{\a}{\alpha}
\renewcommand{\b}{\beta}
\renewcommand{\c}{\gamma}

\def\M{{\cal M}}

\def\be{\begin{equation}}
\def\ee{\end{equation}}
\def\bea{\begin{eqnarray}}
\def\eea{\end{eqnarray}}
\def\ba{\begin{array}}
\def\ea{\end{array}}
\def\bi{\begin{itemize}}
\def\ei{\end{itemize}}

\def\Tr{{\rm Tr}}

\newcommand{\beq}{\begin{equation}}
\newcommand{\eeq}{\end{equation}}
\newcommand{\beqn}{\begin{eqnarray}}
\newcommand{\eeqn}{\end{eqnarray}}
\newcommand{\bga}{\begin{align}}
\newcommand{\w}{\wedge}

\newcommand{\h}{\hspace}
\def\dalemb#1#2{{\vbox{\hrule height .#2pt
\hbox{\vrule width.#2pt height#1pt \kern#1pt
\vrule width.#2pt}
\hrule height.#2pt}}}

\begin{document}

\def\thataddress{Department of Physics, University of Crete, Heraklion 71003, Greece}
\renewcommand\author[1]{#1}

\begin{center}
{\huge { Torsional degrees of freedom in AdS$_4$/CFT$_3$}}
\end{center}
\vskip .8 cm
\centerline{{\bf \author{\large Anastasios C. Petkou\footnote{Based on an invited talk in the 5th Aegean Summer School, "From gravity to thermal field theories: the AdS/CFT correspondence", Adamas, Milos Island, Greece, 21-26 Sep. 2009. The results presented here were obtained in collaboration with R. G. Leigh and N. N. Hoang.}}}}
\vspace{.3cm}
\centerline{\it\thataddress}

\vspace{.5cm}

\begin{abstract}
{} We discuss the holographic implications of torsional degrees of freedom in the context of AdS$_4$/CFT$_3$, emphasizing in particular the physical interpretation of the latter as carriers of the non-trivial gravitational magnetic field, i.e. the part of the magnetic field not determined by the frame field.  As a concrete example we present a new exact 4-dimensional gravitational background with torsion and argue that it corresponds to the holographic dual of a 3d system undergoing parity symmetry breaking. Finally, we compare our new gravitational background with known wormhole solutions - with and without cosmological constant - and argue that they can all be unified under an intriguing "Kalb-Ramond superconductivity" framework.

\end{abstract}

\section{Introduction and summary of the results}

AdS$_4$/CFT$_3$ is currently emerging as a novel paradigm of holography  that has qualitatively different properties from the more familiar AdS$_5$/CFT$_4$ correspondence. Particularly intriguing is the recent accumulation of evidence that AdS$_4$/CFT$_3$ can be used to describe a plethora of phenomena in $2+1$ dimensional systems, such as quantum criticality \cite{QuantCrit1,QuantCrit2}, Quantum Hall transitions \cite{QuantHall1,QuantHall2,QuantHall3,QuantHall4}, superconductivity \cite{Supercond1,Supercond2,Supercond3,Supercond4,Supercond5}, supefluidity \cite{Chrisnew,Basu}  and spontaneous symmetry breaking \cite{SSB1,SSB2,SSB3}.  This has given rise to a whole new research area that goes under the name of AdS/C(ondensed) M(atter) T(heory). Furthermore, AdS$_4$/CFT$_3$ is the appropriate setup to study the holographic consequences of generalized electric-magnetic duality of gravity and higher-spin gauge fields \cite{LP,deHPConf,MPT1,MPT2,deH,Yin}.

In the absence of an explicit AdS$_4$/CFT$_3$ correspondence example\footnote{The recently suggested field theoretic models for M2 branes \cite{BL1,BL2,BL3,Gustavsson,Sangmin,ABJM} are important steps towards the understanding of the boundary side of AdS$_4$/CFT$_3$.} various toy models have been used to study its general qualitative aspects. This work presents yet another model of AdS$_4$/CFT$_3$ which possesses a novel feature. Namely, it can describe the gravity dual of parity symmetry breaking in a 3d system. However, this is not our only aim. We also wish to shed light into torsion from a holographic point of view.  The study of torsion is an interesting subject in itself that poses formal and phenomenological challenges.\footnote{See \cite{Shapiro,Zanelli,Freidel:2005sn} for recent reviews and \cite{Mercuri,Canfora} for other recent works.} In the context of string theory, torsion is omnipresent through antisymmetric tensor fields, therefore AdS$_4$/CFT$_3$ provides the basic setup where it can be holographically investigated. 

This review presents in a slightly expanded form the results of \cite{LNP}. We consider a simple toy model where torsion is introduced via the topological Nieh-Yan class. In particular, we consider the modification of the Einstein-Hilbert action with a negative cosmological constant by the Nieh-Yan class, the latter having a spacetime-dependent coefficient. In the context of the 3+1-split formalism for gravity \cite{LP} we emphasize that the torsional degrees of freedom carry the non-trivial `gravitational magnetic field.' In pure gravity the gravitational magnetic field is fully determined by the frame field and hence  torsion vanishes. In our model, the spacetime dependence of the Nieh-Yan coefficient makes some of the components of the magnetic field dynamical and as a consequence torsional degrees of freedom enter the theory. Our toy model is simple enough such that only one of the torsional degrees of freedom becomes dynamical. This degree of freedom can be either carried by a pseudoscalar, in which case our model is equivalent to a massless pseudoscalar coupled to gravity, or by a two-form gauge potential. In the latter case our model becomes equivalent to a Kalb-Ramond field coupled to gravity.  

Next, we find an exact solution of the equations of motion in Euclidean signature. Our metric ansatz is that of a bulk domain wall (DW). The solution, the {\it torsion DW}, has two distinct asymptotically AdS$_4$ regimes along the ``radial" coordinate. The pseudoscalar has a kink-like profile and it is finite at both of the asymptotic regimes.  Our torsion DW can be viewed as a generalization of the axionic wormhole solution of \cite{GS} in the case of non-zero cosmological constant. See also  \cite{Gutperle} for recent work on AdS wormholes. 

Having in mind the holographic interpretation of our model we focus mainly on the case where the torsional degree of freedom is carried by a pseudoscalar field. Following standard holographic recipes we find that the torsion DW is the gravity dual of a 3d system that possesses two distinct parity breaking vacua. The two vacua are distinguished by the relative sign of the  pseudoscalar order parameter. 
Our bulk picture suggests that the transition from one vacuum to the other can be done by a marginal deformation of the boundary theory. In the Appendix we suggest that the above qualitative properties can be realized in the boundary by the 3d Gross-Neveu model coupled to U(1) gauge fields. 

Further, we point out that the bulk physics of our DW solution bears some intriguing resemblance to the standard Abrikosov vortex in superconducting systems. There is a natural mapping of the parameters of the torsion DW to those of the Abrikosov vortex.  
We show that the gravitational parameter that is interpreted as an order parameter satisfies a $\phi^4$-like equation and this motivates us to suggest that the cosmological constant is related to the ``distance from the critical temperature" as  $\Lambda \sim T_c-T$. However, there is an important difference in that the Abrikosov vortex is  a one-dimensional defect while our DW is codimension one i.e. three-dimensional in AdS$_4$. We also discuss multi-DW configurations and DW condensation and show that $H$-flux supports bubbles of flat spacetime. 

Quite intriguing is our result that DW condensation occurs at a critical value of the magnetic field. This motivates us  to reconsider the known Euclidean solutions of an Einstein-axion system with \cite{Gutperle} and without \cite{GS} cosmological constant. These are wormhole solutions whose salient properties include a quantized electric and (possibly) magnetic flux. Moreover, the $\Lambda\neq 0$ solutions possess a lower bound on their electric flux. Using our intuition that $\Lambda$ plays the role of "temperature", we place the known wormhole and DW solutions on a ("Temperature","Magnetic Field") graph and observe that it resembles a standard superconductivity graph. We call such a system a "Kalb-Ramond superconductor" and we will present more details on its properties in a forthcoming work \cite{LNP2}.


\section{Torsion as the non-trivial magnetic field of gravity}
In this section we discuss the physical interpretation of torsion which is that it carries the non-trivial magnetic degrees of freedom of gravity, namely those that are not determined by the frame field (or, equivalently, by the metric in a 2nd order formulation). To motivate things we recall the first order formalism of electromagnetism in the presence of an $x$-dependent $\theta$-angle, in a non-trivial background here taken to be AdS$_4$. Then we present the 3+1-split formalism for gravity introduced in \cite{LP}. This formalism is a refined form of the standard ADM formalism, which however unveils the physical importance  of the gravitational torsional d.o.f. As we will see, such a point of view is crucial in order to understand the holographic interpretation of torsion.

\subsection{Electromagnetism with a $x$-dependent $\theta$-angle in AdS$_4$}
The vierbeins and metric of AdS$_4$ are 
\beq
e^0=dt\,,\,\,\,e^\a=e^{-t/L}dx^\alpha\,,\,\,\,ds^2=\sigma_\perp dt^2+e^{-2t/L}\eta_{\alpha\beta}dx^\alpha dx^\beta\,,
\eeq
with $\a,\b=1,2,3$. Throughout this work we are being flexible with the both the overall signature and also the nature (spacelilke or timelike)  of the $t$-direction i.e.we set $\eta_{\alpha\beta}={\rm diag}(1,1,\sigma_3)$, $\sigma_\perp\sigma_3=\sigma=\pm1$. The gauge potential and field strength are one-forms 
 \beq
 A=A_0dt+\tilde{A}\,,\,\,\,\,\,\,\, F=-dt\wedge E+\tilde{F}=\frac{1}{2}F_{ab}e^a\wedge e^b\,,
 \eeq
 where $a,b=0,1,2,3$ and $E=-F_{0\a}e^\a$ is the electric field. The tilde will always denote quantities along the three directions $1,2,3$. With the above definitions we find
 \beq
 *_4F=dt\wedge *_3\tilde{F} -\sigma_\perp *_3E\,,\,\,\, dA=dt\wedge(\dot{\tilde{A}}-\tilde{d}A_0)+\tilde{d}\tilde{A}\,.
 \eeq
Recall e.g. that $*_3e^i=\frac{1}{2}\epsilon^{i}_{\phantom{i}jk}e^j\wedge e^k$ and $*_3e^i\wedge e^k=\epsilon^{ij}_{\phantom{ij}k}e^k$. Note also that $*_4^2=\sigma_\perp$, $*_3^2=\sigma_3$ and $\epsilon_{0123}=1$. 
 The first order action is
 \beq
 \label{1storderEM}
 I=\int -dA\wedge *_4F+\frac{1}{2}F\wedge *_4F+\frac{\theta}{2}dA\wedge dA\,.
 \eeq
Notice that due to the $x$-dependance of $\theta$ the last term in (\ref{1storderEM}) is not a total derivative and will give contributions to the e.o.m. After some work the action above takes the more familiar form
\beqn
I&=&\sigma_\perp\int dt\wedge\left[\dot{\tilde{A}}\wedge\left( *_3 E+\sigma_\perp\theta\tilde{d}\tilde{A}\right)+\frac{1}{2}\left(E\wedge *_3E-\sigma_\perp\tilde{F}\wedge *_3\tilde{F}\right) \right.\nonumber \\
&&\hspace{2cm}+A_0\left(\tilde{d}*_3E+\sigma_\perp\tilde{d}\theta\wedge \tilde{d}\tilde{A}\right)\Bigl]\,.
\eeqn
This gives the Hamiltonian e.o.m.
\beq
E=-\dot{\tilde{A}} \,,\,\,\,\,\tilde{d}B=-\sigma_\perp\dot{\left(*_3E\right)}-\dot{\theta}\sigma_3*_3B-\tilde{d}\theta\wedge E\,,
\eeq
where we have defined the magnetic field (also a one-form) as
\beq
\tilde{F}=\tilde{d}\tilde{A}\equiv \sigma_3(*_3B)\,.
\eeq
We also have the Gauss law and Bianchi identity respectively
\beq
\tilde{d}*_3E+\sigma_\perp\tilde{d}\theta\wedge\tilde{d}\tilde{A}=0\,,\,\,\,\,\,\,\tilde{d}*_3B=0\,.
\eeq
It is straightforward to show that the above give the Maxwell equations in the more familiar form
\beqn
&&\vec{\nabla}\times\vec{E}=-\sigma_3\frac{\partial\vec{B}}{\partial t}\,,\,\,\,\,\,\,\vec{\nabla}\cdot\vec{E}=-\sigma_\perp\sigma_3\vec{\nabla}\theta\cdot\vec{B}\,,\\
&& \vec{\nabla}\times\vec{B}=-\sigma\frac{\partial\vec{E}}{\partial t}-\left[\dot{\theta}\vec{B}+\vec{\nabla}\theta\times\vec{E}\right] \,,\,\,\,\,\,\,\vec{\nabla}\cdot\vec{B}=0\,.
\eeqn
We summarize the effects of an $x$-dependent $\theta$-angle in electromagnetism:
\begin{itemize}
\item The modification of the canonical momentum as we see in (5)
\beq
*_3E\mapsto *_3E+\sigma\theta *_3B\,.
\eeq
\item The presence of a source term for Gauss law. 
\end{itemize}
In particular, there is no additional d.o.f. introduced by the $\theta$-angle. We will compare this situation with gravity in the following.

\section{Details on the the 3+1-split formalism}
In this section we present a concise version of the 3+1-split formalism of \cite{LP} for  gravity in the presence of non-zero cosmological constant. We consider a globally hyperbolic Lorentzian manifold ${\cal M}$ and take the Einstein-Hilbert action with cosmological constant in the first-order Palatini formalism as
\beq\label{EHP}
S_{{\rm EH}}=-\frac1{32\pi G}\int_{\M}\epsilon_{abcd}\left(R^{ab}+\frac\Lambda2e^a\wedge e^b\right)\wedge e^c\wedge e^d\,.
\eeq
This is thus equivalent to the standard second-order gravitational action 
\beq
S_{{\rm2^{nd}}}=-\frac1{16\pi G}\int d^{4}x\sqrt{-g}\left(R+6\Lambda\right)\,,
\eeq
and hence the cosmological constants is related to the parameter $\Lambda$ as $\Lambda_{\rm cosm.}=-3\Lambda$.
The curvature and torsion 2-forms are defined in terms of the vielbein $e^a$  and spin-connection $\omega^{ab}$ as
\beq
R^{ab}=d\omega^{ab}+\omega^a{}_{c}\wedge\omega^{cb}\,,\qquad T^a=d e^a+\omega^a{}_{b}\wedge e^b\,.
\eeq
We define as before $\eta_{ab}={\rm diag}(\sigma_{\perp},+,+,\sigma_3)$, where $\sigma_{\perp}\sigma_3=\sigma=\pm 1$, $\sigma_\perp^2=\sigma_3^2=1$ and set $\Lambda=\sigma_{\perp}/\ell^{2}$ such that $\Lambda <0$ ($\Lambda >0$) yields the de Sitter (Anti-de Sitter) vacuum.  Next, we split the vielbein and the spin connection as
\beqn
e^0&=&Nd t\,,\qquad e^\alpha=N^\alpha d t+\tilde e^\alpha\,,\label{esplit}\\
\omega^{0\alpha}&=&q^{0\alpha}d t+\sigma_{\perp}K^\alpha\,,\qquad\omega^{\alpha\beta}=-\epsilon^{\alpha\beta\gamma}\left(Q_\gamma d t+B_\gamma\right)\,.\label{omegasplit}
\eeqn
The novelty of the formalism is the introduction of the gravitational electric $K^\alpha$ and magnetic fields $B^\alpha$, which are both vector-valued one-forms on the slices. 
We then find for the torsion
\beq T^\alpha =\tilde T^\alpha+dt\wedge\left\{\dot{\tilde e}^\alpha-\tilde d N^\alpha+NK^\alpha-\sigma{\epsilon^\alpha}_{\beta\gamma}Q^\beta e^\gamma-\sigma{\epsilon^\alpha}_{\beta\gamma}N^\beta B^\gamma\right\}\,,
\eeq
\beq \label{T0}
T^0 =\sigma_\perp K_\alpha\wedge\tilde e^\alpha+dt\wedge\left\{-\tilde d N-\sigma_\perp N_\alpha K^\alpha+{q^0}_\beta\tilde e^\beta\right\}\,,
\eeq
and we write
\beq
{R^a}_b=\tilde R^{a}_{\kern5pt b}+dt\w r^{a}_{\kern5pt b}\,,
\eeq
\begin{equation}\label{eq:GCtwo2}
{\tilde R^0}_{\kern5pt\alpha}=\sigma_\perp ( {\tilde {d}K}_\alpha+K_\beta\wedge\tilde\omega^\beta_{\kern5pt\alpha})\equiv\sigma_\perp (\tilde D K)_\alpha\,,
\end{equation}
\begin{equation}\label{eq:GCone1}
\tilde R^{\alpha}_{\kern5pt\beta}=^{(3)}\kern-5pt{R^{\alpha}}_{\beta}- \sigma_\perp K^\alpha\wedge K_\beta\,,
\end{equation}
with
\beq
^{(3)}R^{\alpha\beta}=
\sigma \left[\epsilon^{\alpha\beta\gamma} dB_\gamma-\sigma_\perp B^\alpha\w  B^\beta\right]\,,
\eeq
and
\begin{eqnarray}
2\epsilon_{\alpha\beta\gamma}r^{0\alpha}\wedge {\tilde e}^\beta\wedge {\tilde e}^\gamma &=&
 2\sigma_\perp\epsilon_{\alpha\beta\gamma} \dot K^\alpha\wedge\tilde e^\beta\wedge\tilde e^\gamma+4Q_\alpha K_\beta\wedge\tilde e^\beta\wedge\tilde e^\alpha\nonumber \\
 &&+4q^{0\alpha}\left[\epsilon_{\alpha\beta\gamma}\tilde{T}^\beta\wedge\tilde{e}^\gamma\right]\,.
\end{eqnarray}
After some tedious but straightforward calculations we find
\beqn
\label{SEH1}
S_{EH}&=&-\frac{\sigma_{\perp}}{8\pi G}\int dt\wedge \Bigl\{ \dot{\tilde e}^\alpha\wedge (4\sigma_\perp\epsilon_{\alpha\beta\gamma}K^\gamma\wedge\tilde e^\beta)-4\sigma_\perp N^\alpha\epsilon_{\alpha\beta\gamma}(\tilde D K)^\beta\wedge {\tilde e}^\gamma
\nonumber \\
&&\hspace{+1cm}+ 2N\epsilon_{\alpha\beta\gamma}\left( ^{(3)}R^{\alpha\beta}-\sigma_\perp K^\alpha\wedge K^\beta-\frac{\Lambda}{3} {\tilde e}^\alpha\wedge {\tilde e}^\beta\right)\wedge {\tilde e}^\gamma \nonumber \\ &&\hspace{+1cm}  -4q^{0\alpha}\epsilon_{\alpha\beta\gamma}\tilde T^\beta\w \tilde e^\gamma+4Q_\alpha\tilde e^\alpha\w K_\beta\w\tilde e^\beta\Bigl\}-S_{GH}
\eeqn
where the last term is exactly the usual Gibbons-Hawking surface term
\beq
S_{GH} =-\frac1{16\pi G}\int_{\partial\M}\left(q^{0\alpha}d t+\sigma_{\perp}K^{\alpha}\right)\wedge\epsilon_{\a\b\c}\tilde{e}^\b\wedge\tilde{e}^c\,.
\eeq
{\it Adding} then the Gibbons-Hawking term in (\ref{SEH1}) we obtain
\beqn
\hat{S}_{EH}&=&-\frac{\sigma_{\perp}}{8\pi G}\int_{\M}d t\wedge\left\{-K_{\alpha}\wedge\dot\Sigma^\alpha+N\tilde W_{\alpha}\wedge\tilde e^{\alpha}+\sigma_\perp\hat Q\wedge K_{\beta}\wedge\tilde e^{\beta}\right.\nonumber\\
&&\qquad\qquad\qquad\left.+\sigma_{\perp}q^{0\alpha}\tilde{D}\Sigma_{\alpha}-N^{\alpha}\epsilon_{\alpha\beta\gamma}\tilde{D}K^{\beta}\wedge\tilde e^{\gamma}\right\}\,,\label{lor.act}
\eeqn
where $\hat Q\equiv Q_{\alpha}\tilde e^{\alpha}$. We have introduced the 2-form
\beq
\tilde W_{\alpha}\equiv\rho_{\alpha}-\frac12\epsilon_{\alpha\beta\gamma}K^{\beta}\wedge K^{\gamma}+\frac1{\ell^{2}}\Sigma_{\alpha}\,.\label{tildeW}
\eeq
and have defined the oriented surface element as
\beq
\Sigma^{\alpha}=*_3\tilde e^{\alpha}=\frac12\epsilon^{\alpha}{}_{\beta\gamma}\tilde e^{\beta}\wedge\tilde e^{\gamma}\,,
\eeq
with $*_3$ the three-dimensional Hodge dual defined in terms of $\tilde e^\alpha$ only. The three-dimensional component of the curvature 2-form 
\beq
\rho_{\alpha}=\tilde d B_{\alpha}+\frac12\epsilon_{\alpha\beta\gamma}B^{\beta}\wedge B^{\gamma}\,,
\eeq
is made out of $B^\alpha$ only. Recall (i.e. (19)) that $\tilde{D}$ is a covariant derivative with respect to the one-form field $B^{\alpha}$ as 
\beq
\tilde{D}V^\alpha =\tilde d V^\alpha +\epsilon^{\alpha}_{\,\,\beta\gamma}B^\beta\wedge V^\gamma\,,
\eeq
if $V^\alpha$ is a generic vector-valued one-form (with respect to either ${\rm SO}(3)$ or ${\rm SO}(2,1)$ depending on whether $\sigma_\perp=\mp1$ respectively) defined on $\Sigma_t$. Comparing the action (\ref{lor.act}) to the electromagnetic action (\ref{1storderEM}) motivates calling the vector-valued  one-forms $K^\alpha$ and $B^\alpha$ the ``electric" and ``magnetic"  fields respectively. 

The action (\ref{lor.act})  is stationary on-shell when $\delta\tilde{e}^\alpha=0$ in the boundary, i.~e.~it provides a good Dirichlet variational principle with respect to the vielbein. The form of the action (\ref{lor.act}) appears to indicate that the proper conjugate dynamical variables are $\Sigma^{\alpha}$ (or, equivalently, $\tilde e^\alpha$) and $K^\alpha$. It has been shown in \cite{MPT1,MPT2} that the proper identification of the dynamical variables is slightly more involved than this.  The remaining fields  $\{N,N^\alpha,q^{0\alpha},\hat Q,B^{\alpha}\}$ enter the action as Lagrange multipliers of the following constraints:
\beqn
-8\pi G\sigma_{\perp}\frac{\delta S}{\delta N}&=&\tilde W_{\alpha}\wedge\tilde e^{\alpha}=0\,,\label{lC.1}\\
-8\pi G\sigma_{\perp}\frac{\delta S}{\delta N^\alpha}&=&-\epsilon_{\alpha\beta\gamma}\tilde{D}K^\beta\wedge\tilde e^\gamma=0\,,\label{lC.2}\\
-8\pi G\sigma_{\perp}\frac{\delta S}{\delta q^{0\alpha}}&=&\sigma_{\perp}\tilde{D}\Sigma_{\alpha}=\sigma_{\perp}\epsilon_{\alpha\beta\gamma}\tilde T^{\beta}\wedge\tilde e^{\gamma}=0\,,\label{lC.3}\\
-8\pi G\sigma_{\perp}\frac{\delta S}{\delta\hat Q}&=&\sigma_\perp K_\alpha\wedge\tilde e^\alpha=0\,,\label{lC.4}\\
-8\pi G\sigma_{\perp}\frac{\delta S}{\delta B^{\alpha}}&=&N\tilde T^{\alpha}+\left(\tilde d N+\sigma_{\perp}K_{\beta}N^{\beta}-\hat q\right)\wedge\tilde e^{\alpha}=0\,, \label{lC.5}
\eeqn
where $\hat q\equiv q^{0}{}_{\alpha}\tilde e^{\alpha}$. The exterior multiplication of (\ref{lC.5}) by $\epsilon_{\alpha\beta\gamma}\tilde{e}^\gamma$ gives, by virtue of (\ref{lC.3}),
\beq
\tilde d N+\sigma_{\perp}K_{\beta}N^{\beta}-\hat q=0\,,\label{lC.6}
\eeq
and hence we obtain the zero torsion condition
\beq
\label{torsion0}
\tilde T^{\alpha}=\tilde{D}\tilde{e}^\alpha=0\,,
\eeq
The last equation unveils the physical meaning of the gravitational magnetic field $B^\a$: it is a Lagrange multiplier which is algebraically related to the vielbein via the vanishing of torsion (\ref{torsion0}). This is exactly analogous to electromagnetism and gives an important hint regarding the relevance of torsion to holography and gravitational duality \cite{LNP}.

\subsection{The analog of $\theta$-angle in gravity: the Nieh-Yan invariant}

There is a number of topological terms built from the gravitational dynamical variables that one may consider in 4 dimension. These are all of potential interest to holography  because being total derivatives  they may induce interesting boundary effects. We may parameterize these terms as follows (writing all possible $SO(3,1)$-invariant 4-forms constructed from $e^a, {R^a}_b, T^a$):
\beq
I_{top}=n\int_M C_{NY}+2\gamma^{-1}\int_M C_{Im}+p\int_M P_4+q\int_M E_4\,,
\eeq
where 
\beq C_{NY}= T^a\wedge T_a-R_{ab}\wedge e^a\wedge  e^b=d(T^a\wedge e_a)\,,
\eeq
is the Nieh-Yan form. The parameter $\gamma$ is often referred to as the Immirzi parameter with 
\beq
C_{Im}={R^a}_b\wedge e^b\wedge e_a\,.
\eeq
 The remaining objects are the Pontryagin form
 \beq
 P_4=-\frac{1}{8\pi^2}{R^a}_b\w {R^b}_a=-\frac{1}{8\pi^2}d({\omega^a}_b\wedge {R^b}_a-\frac13 {\omega^a}_b\w {\omega^b}_c\w {\omega^c}_a)\,,
 \eeq
 and the Euler form
 \beq
 E_4=-\frac{1}{32\pi^2}\epsilon_{abcd}R^{ab}\w R^{cd}\,.
 \eeq
We note that $P_4+\frac{\sigma_\perp a^2}{4\pi^2} C_{NY}$ and $C_{NY}-C_{Im}$ are actually $SO(3,2)$ invariants \cite{Zanelli}.
These terms become very interesting even in gravity if we allow the coefficients to become fields (for some interesting recent literature on the matter see \cite{Mercuri,Canfora}).


\subsection{Torsion and the magnetic field of gravity}

In pure Eistein-Hilbert gravity the torsional d.o.f. are not dynamical and are carried by the magnetic field $B^\alpha$ . This is seen for example if we recall the definition of  the non-trivial `spatial' torsion as
\beq
\label{3torsion}
\tilde{T}^\alpha =\tilde{d}\tilde{e}^\alpha -\sigma\epsilon^{\alpha\beta\gamma}B_\beta\wedge\tilde{e}_\gamma\,.
\eeq
Moreover, it is  seen from (\ref{T0}) that the radial component of torsion $T^0$ is determined by $\tilde{e}^\alpha$ and $K^\alpha$. Notice that (\ref{3torsion}) implies that the tensor $B_{\alpha\beta}$ is odd under `spatial' parity, hence its trace $B^\alpha_{\,\,\alpha}$ is a pseudoscalar. Hence, although a priori the torsional degrees of freedom are not connected with the pair of conjugate variables $\tilde{e}^\alpha $ and $K^\alpha$, they are not dynamical as there is no kinetic term for $B^\alpha$. Rather, they enter (\ref{lor.act}) algebraically and as such they give via (\ref{lC.5}) and (\ref{lC.6}) the algebraic zero torsion condition (\ref{torsion0}) by virtue of which the magnetic field is related to the frame field. This is the gravitational analogue of the electromagnetic case where the magnetic field is related to the gauge potential via the Bianchi identity. 

Consider now adding to the Einstein-Hilbert action the Nieh-Yan class $C_{NY}$ with a constant coefficient $\theta$. Over a compact manifold, the NY class is a topological invariant and takes integer values\footnote{More precisely, $C_{NY}/(2\pi L)^2$ is integral, as it is equal to the difference of two Pontryagin forms, one $SO(3,2)$ and one $SO(3,1)$.} \cite{Zanelli}. Having in mind holography, we are interested here in manifolds with  boundary. In particular, the $3+1$ split has been set up so that the boundary is a constant-$t$ slice.  The NY term reduces to a boundary contribution. The explicit calculation yields
\beqn
\label{NY3+1}
{\cal I}_{NY} \equiv -2\sigma_\perp\theta\int C_{NY}
= 2\sigma_\perp \theta\int dt\wedge \left[2\epsilon_{\alpha\beta\gamma}\dot{\tilde{e}}^\a\wedge\tilde{e}^\beta\wedge B^\gamma +\epsilon_{\alpha\beta\gamma}\dot{B}^\alpha\wedge\tilde{e}^\beta\wedge\tilde{e}^\gamma\right]
\eeqn
Adding (\ref{NY3+1}) to (26) we obtain
\beqn
\label{3+1EHNY}
\hat{S}_{EH}+
{\cal I}_{NY}&\propto & \int dt\wedge\Bigl(\dot{\tilde e}^\alpha\wedge (-4\sigma_\perp\epsilon_{\alpha\beta\gamma}\tilde{e}^\beta\wedge \left[ K^\gamma-\theta B^\gamma\right])\nonumber \\
&&+2\sigma_\perp\theta\epsilon_{\alpha\beta\gamma}\dot{B}^\alpha\wedge\tilde{e}^\beta\wedge\tilde{e}^\gamma +\ {\rm constraint\ terms}\Bigl) \,.
 \eeqn
Notice that the ${\cal I}_{NY}$ term has two effects. One is to modify the canonical momentum variable $K^\alpha\mapsto K^\alpha -\theta B^\alpha$. This is analogous to the effect of the $\theta$-angle in the canonical description of electromagnetism in section 2.1.  The other is to 
provide a kinetic term for the singlet component of the magnetic field (one easily verifies that only ${B^\alpha}_{\alpha}$ contributes in the second term in the first line of (\ref{3+1EHNY})). This second effect has no analogue in electromagnetism. Taking the variation of (\ref{3+1EHNY}) with respect to $B^\alpha$, one finds that the zero torsion condition still holds.  This is expected of course since the ${\cal I}_{NY}$  term is  purely a boundary term. As a consequence, the true dynamical variables remain $\tilde{e}^\alpha$ and $K^\alpha$. However, the holography is slightly modified. The variation of (\ref{3+1EHNY}) gives on-shell 
\beq
\label{NYHol1}
\delta\left(\hat{S}_{EH}+{\cal I}_{NY}\right)_{on\,\,shell} \propto \int_{\partial {\cal M}}\delta\tilde{e}^\alpha\wedge\left(-4\sigma_\perp\epsilon_{\alpha\beta\gamma}\tilde{e}^\beta\wedge\left[K^\gamma-\theta B^\gamma\right]\right)_{on\,\,shell}\,.
\eeq
After the appropriate subtraction of divergences \cite{MPT1,MPT2}, (\ref{NYHol1}) yields a modified boundary energy momentum tensor. The modification is due to the term $4\sigma_\perp \theta\epsilon_{\alpha\beta\gamma}\tilde{e}^\beta\wedge B^\gamma$ which is parity odd and corresponds to the unique symmetric, conserved and traceless tensor of rank two and scaling dimension three that can be constructed from the three-dimensional metric \cite{LPsl2z}
\beq
T_{\a\b}^{bdry} \mapsto T_{\a\b}^{bdry} +\theta T_{\a\b}^{top}\,,\,\,\,\, \,\,\,\,T_{\a\b}^{top}\propto \epsilon_{\ell m(\a}\partial_\ell\partial^2 g_{\b)m}
\eeq
where $g_{\a\b}$ being the boundary metric. It is  the exact analogue of the topological spin-1 current constructed from the 3d gauge potential \cite{Witten,LPsl2z}.

The form of the action (\ref{3+1EHNY}) unveils an intriguing possibility. The above holographic interpretation was based on the zero torsion condition that connects $B^\alpha$ to the frame field. However, to get the zero torsion condition from (\ref{3+1EHNY}) we needed to integrate by parts the last term in the first line. Hence, if $\theta$ were $t$-dependent, the torsion would no longer be zero and the trace $B^\alpha_{\,\,\alpha}$ would become a proper dynamical degree of freedom independent of $\tilde{e}^\alpha$. In such a case the holographic interpretation of (\ref{3+1EHNY}) would change. The new bulk degree of freedom would couple to a new pseudoscalar boundary operator. As a consequence, we have the possibility to probe additional aspects of the boundary physics and describe new 2+1 dimensional phenomena. That we do in the next section. 

\section{The Nieh-Yan models}
\subsection{General aspects}

In the previous section we sketched a mechanism by which torsional degrees of freedom become dynamical. In particular, we have argued that the addition of the Nieh-Yan class with a space-time-dependent coefficient in the Einstein-Hilbert action makes dynamical one pseudoscalar degree of freedom which is connected to the trace of the gravitational magnetic field. 
Adding boundary terms to the bulk action corresponds to a canonical transformation. Consequently, by adding boundary terms we can change the canonical interpretation  and the variational principle.  Consider first the action
\beq\label{eq:INYprime}
I'_{NY}=\hat{S}_{EH}[e,\omega]+I_{GH}[e,\omega]+2\int_M F(x) C_{NY}\,,
\eeq
where $F$ is a pseudoscalar `axion' field with no kinetic term. If $F\equiv -\sigma_\perp\theta$ were a constant, this theory would be equivalent to that studied in the last section. With $F=F(x)$, we have additional terms in the action involving gradients of $F$. If we perform the $3+1$ split on this action, we will find that $\tilde e^\alpha$ and $B^\alpha$ are canonical coordinates, and their conjugate momenta will depend on $F$.


 
The action as given may be supplemented by additional boundary terms. Such boundary terms are analogous to the Gibbons-Hawking term in gravity, but here involve the torsional degrees of freedom. In particular, we can replace $I'_{NY}$ by 
\beq
I_{NY}=\hat{S}_{EH}[e,\omega]+I_{GH}[e,\omega]-2\int_M dF\w T_a\w e^a\,.
\eeq
This action is such that $\tilde e^\alpha$ and $F$ are canonical coordinates with appropriate boundary conditions, while $B^\alpha$ appears in the  momentum conjugate to $F$. 
To investigate this theory, we note that the variation of the action takes the form
\beqn
\delta I_{NY}=2\int_M \delta e^d\w  \left[\epsilon_{abcd} e^b\w\left(R^{cd}-\frac13\Lambda e^c\w e^d\right)+2dF\w T_d\right]\nonumber \\
+2\int_M\delta\omega^{ab}\w\left[ \epsilon_{abcd}T^c\w e^d+ dF\w e_b\w e_a
\right]+2\int_M\delta F\ C_{NY}\nonumber \\
+2\int_Md[\delta e^a\wedge\left(\epsilon_{abcd} e^b\w\omega^{cd}-dF\w e_a\right)-T_a\w e^a\delta F ]\,.
\eeqn
A non-trivial configuration of $F$ sources a particular component of the torsion. Indeed the classical equations of motion can be manipulated to yield in the bulk
\beq\label{eq:Ftors}
T^a\w e_a=3*_4 dF\,,
\eeq
where $*_4$ denotes the Hodge-$*$ operation.
However, as d'Auria and Regge \cite{dr} showed, this classical system is equivalent to a pseudoscalar
coupled to torsionless gravity. 
\beq\label{eq:PSL}
I_{PS}=\hat{S}_{EH}[e,\cc{\omega}]+I_{GH}[e,\cc{\omega}]-3\int_M dF\w *_4 dF\,.
\eeq
This comes about as follows. We write the connection as $\omega=\cc{\omega}+\Omega$, where $\cc{\omega}$ is torsionless, and insert the
equation of motion (\ref{eq:Ftors}). The latter becomes an equation\footnote{Explicitly this is ${\Omega^a}_b=\frac{\sigma}{4}{\epsilon^{acd}}_b \pa_c F e_d$.} for $\Omega$, and we obtain (\ref{eq:PSL}). 

A massless pseudoscalar field coupled to torsionless gravity is holographically dual to composite pseudoscalar operators of dimensions $\Delta=3,0$ in the boundary. The usual holographic dictionary then says that only the $\Delta=3$ operator appears in the boundary theory since only this is above the unitarity bound of the 3d conformal group $SO(3,2)$. A scalar operator with  dimension $\Delta=0$ would simply correspond to a constant in the boundary. Hence, the sensible holographic interpretation of the massless bulk pseudoscalar is that its leading behaviour determines the marginal coupling of a $\Delta=3$ operator; the expectation value of the operator itself is determined by the subleading behaviour of the bulk pseudoscalar. 

Another equivalent formulation of this bulk theory is obtained by writing
\beq
*_4dF={\frac{1}{3}} H\,.
\eeq
with $H$ a 3-form field.
 This is the parameterization that would be most familiar from string theory, 
as the system simply corresponds to an antisymmetric 2-form field. In this formulation, we write
\beqn
I_{KR}&=&\hat{S}_{EH}[e,\cc{\omega}]+I_{GH}[e,\cc{\omega}]+\frac13\int_M H\w *_4H+\sqrt{\frac{2}{3}}\int_MC\w d*_4H\nonumber \\
&=&\hat{S}_{EH}[e,\cc{\omega}]+I_{GH}[e,\cc{\omega}]-\frac12\int_M dC\wedge *_4dC+\int_M d(C\w *_4dC)\,.
\eeqn
In the first equation, $C$ appears as a Lagrange multiplier for the `Gauss constraint' and in the second expression, we have solved for the $H$ equation of motion in the bulk, which is just $H=\sqrt{\frac32}dC$.

\subsection{The 3+1-split of the pseudoscalar Nieh-Yan model}

To investigate the holographic aspects of our model it is most useful to use the `radial quantization' in which we think of the radial coordinate as `time' $t$. The Nieh-Yan deformation gives
\beqn
-2\int dF\wedge T^a\wedge e_a &=& 2\int dt\w \Biggl\{-\dot F\tilde T_\alpha\w \tilde e^\alpha-\dot{\tilde e}_\alpha\w\tilde d F\w\tilde e^\alpha +N[2\tilde dF\w K_\alpha\w\tilde e^\alpha]\nonumber \\
 && \hspace{1cm}+N^\alpha[2\tilde dF\w\tilde T_\alpha]+Q^\alpha[-\sigma\epsilon_{\alpha\beta\gamma}\tilde dF\w\tilde e^\beta\w \tilde e^\gamma]\Biggl\}\,.
\eeqn
We see that the $F$ field makes a contribution to the constraints, and has a conjugate momentum proportional to the scalar part of the torsion (the part transverse to the radial direction).
The full bulk action becomes
\beqn
I&=&\int dt\wedge\left(\dot{\tilde e}^\alpha\wedge (4\sigma_\perp\epsilon_{\alpha\beta\gamma}K^\gamma\wedge\tilde e^\beta-2\tilde dF\wedge\tilde e_\alpha)-2\dot F (\tilde e^\alpha\wedge\tilde T_\alpha)\right.\nonumber \\
&&+N\left\{ 2\epsilon_{\alpha\beta\gamma}\left( ^{(3)}R^{\alpha\beta}-\sigma_\perp K^\alpha\wedge K^\beta-\frac{\Lambda}{3} {\tilde e}^\alpha\wedge {\tilde e}^\beta\right)\wedge {\tilde e}^\gamma 
+4\tilde dF\wedge K_\alpha\wedge\tilde e^\alpha
\right\}\nonumber\\
&&+4{N}^\alpha\left\{-\sigma_\perp \epsilon_{\alpha\beta\gamma}(\tilde D K)^\beta\wedge {\tilde e}^\gamma+\tilde dF\wedge\tilde T_\alpha
\right\}\nonumber \\
&&+4Q^\alpha\left\{ (K_\beta \wedge\tilde e^\beta)\wedge\tilde e_\alpha-\frac12\sigma\epsilon_{\alpha\beta\gamma}
\tilde dF\wedge\tilde e^\beta\wedge\tilde e^\gamma
\right\}\nonumber \\
&&+\left.4{q^0}_\alpha\left\{
{\epsilon^\alpha}_{\beta\gamma}\tilde{T}^\beta\wedge\tilde{e}^\gamma
\right\}\right)\,.
 \eeqn
We notice that the $Q$-constraint term can be written in the form
\beq\label{eq:Kcon}
4Q_\alpha\tilde e^\alpha\w\left( K_\beta \wedge\tilde e^\beta-\sigma *_3\tilde dF\right)\,.
\eeq
Because of this constraint (which relates the antisymmetric part of the extrinsic curvature to
the vorticity of $F$), the momentum conjugate to $\tilde e^\alpha$ is symmetric, i.e.
\beqn
\Pi_\alpha&=& 4\sigma_\perp\epsilon_{\alpha\beta\gamma}K^\gamma\wedge\tilde e^\beta-2\tilde dF\wedge\tilde e_\alpha\nonumber \\
&=& 4\sigma_\perp\left(\epsilon_{\alpha\beta\gamma}K^\gamma\wedge\tilde e^\beta-\frac12\sigma_3 *_3(K_\beta\w \tilde e^\beta)\w \tilde e_\alpha\right)\,.
\eeqn
When written out in components, one finds that the antisymmetric part $K_{[\alpha\beta]}$ cancels
\beq
\Pi_\alpha = 4\sigma_\perp (K_{(\beta\alpha)}-tr K\ \eta_{\beta\alpha})\tilde e^\beta\,.
\eeq
This result is consistent with the fact noted above, that the system may be equivalently described as
a pseudoscalar field coupled to torsionless gravity. Moreover, if  we take the deDonder gauge $d^\dagger\tilde e^\alpha=0$,  the torsion constraint implies that $B$ is symmetric. 

The ${q^0}_\alpha$ constraint yields ${\tilde T^\beta}_{\alpha\beta}=0$.
Out of the nine components of $\tilde T$, which transform as ${\bf 5}+{\bf 3}+{\bf 1}$ under $SO(3)$ (or $SO(2,1)$), this
sets the triplet to zero (the ${\bf 5}$ also vanishes on an equation of motion). The momentum conjugate to $F$ is given by
\beq
\Pi_F = -2\epsilon^{\alpha\beta\gamma}\tilde T_{\alpha\beta\gamma}\,.
\eeq
This is the singlet part of the torsion, which has become dynamical in this description of the theory, in the sense that it is canonically conjugate to $F$.

\section{The torsion domain wall}

We will now simplify the analysis by taking a coordinate basis and looking for solutions of the form
\beq
\tilde e^\alpha = e^{A(t)}dx^\alpha,\ \ \ \ N=1,\ \ \ \ N^\alpha=0\,,
\eeq
and we will further suppose that $F=F(t)$. In this case $K^\alpha$ and $B^\alpha$ reduce to one degree of freedom each as a result of the constraints
\beq
K_\alpha=k\tilde e_\alpha,\ \ \ \  B_\alpha=b\tilde e_\alpha\,,
\eeq
and one finds $\Pi_A=-4\sigma_\perp k$ and $\Pi_F=2\sigma b$. The action then takes the following relatively simple Hamiltonian form
\beq
I_{NY} \propto \int dt\,d^3x\,\,e^{3A(t)}\left[ \dot A\Pi_A+\dot F\Pi_F-\left(\frac12\sigma_3\Pi_F^2+\frac18\sigma_\perp\Pi_A^2+\frac23\Lambda\right)\right]\,.
\eeq
and the equations of motion give
\beqn
\label{eom1234}
&\dot\Pi_A=3\dot F\Pi_F,\ \ \ \ \ \ \dot\Pi_F+3\Pi_F\dot A=0,\ \ \ \ \ \
\Pi_A=4\sigma_\perp\dot A,\ \ \ \ \ \Pi_F=\sigma_3\dot F&\,,\\ 
\label{eom5}
&\Pi_A^2+4\sigma\Pi_F^2+\frac{16}{3}\sigma_\perp\Lambda=0&\,.
\eeqn
These equations of motion could of course alternatively  be obtained by considering
the theory in the form (\ref{eq:PSL}). It is convenient to rescale $F(t)=\frac13 \Theta(t)$. Then the equations of motion can be put in the form
\beq
\label{dwalleqs}
\dot{A}+3\dot A^2-3a^2=0, \ \ \ \ \ \ \dot{A}=\frac{1}{12}\sigma\dot \Theta^2, \ \ \ \ \ \ \dot{\Theta}+3\dot \Theta\dot A=0\,.
\eeq
where we have set $\Lambda=-3\sigma_\perp a^2$ with $a=1/L$. These are of the standard form of domain wall equations that have appeared numerous times in the AdS/CFT literature. However, there is a crucial difference. Notice that the first two of (\ref{dwalleqs}) imply
\beq
\label{dwalleq1}
\dot{A}^2 +\frac{1}{36}\sigma\dot{\Theta}^2 -a^2=0\,.
\eeq
For Euclidean signature ($\sigma=\sigma_3=1$) the second term in (\ref{dwalleq1}) has {\it positive} sign in contrast to most of the other holographic studies. This is due to the fact that in passing from Lorentzian to Euclidean signature the pseudoscalar kinetic term acquires the `wrong sign' \cite{Gibbons}. This property allows for a remarkable exact solution to the above  system of non-linear equations in Euclidean signature, which we refer to as the {\it torsion DW}.  To obtain it we define
\beq
h(t)=\dot A(t)\,,
\eeq
at which point we have
\beq\label{eq:hdot}
\dot h=\frac{1}{12}\dot \Theta^2,\ \ \ \dot h+3(h^2-a^2)=0\,.
\eeq
The general solution is of the form
\beq
h(t)=a\tanh 3a(t-t_0)
\eeq
and we then have
\beq
\label{Pf}
\Pi_F=\dot F=\pm2\sqrt{a^2-h^2(t)}=\pm 2a\ \text{sech}\ 3a(t-t_0)
\eeq
which gives
\beq
\label{Theta}
\Theta(t)=\Theta_0\pm 4\arctan\left(e^{3a(t-t_0)}\right)\,.
\eeq
The $\pm$ sign corresponds to kink/antikink and we will without loss of generality choose the $+$ sign.
We may also solve for
\beq\label{eq:expA}
e^{A(t)}=\a (2\cosh 3a(t-t_0))^{1/3}
\eeq
The parameter $\alpha$ is an arbitrary positive integration constant that sets the overall scale of the spatial part of the metric. $t_0$ may be interpreted as the position of the DW; when $t_0=0$ the torsion DW sits in the middle between the two asymptotically AdS$_4$ regimes.  
Below, we will discuss the interesting holographic interpretation of the torsion DW.

\myfig{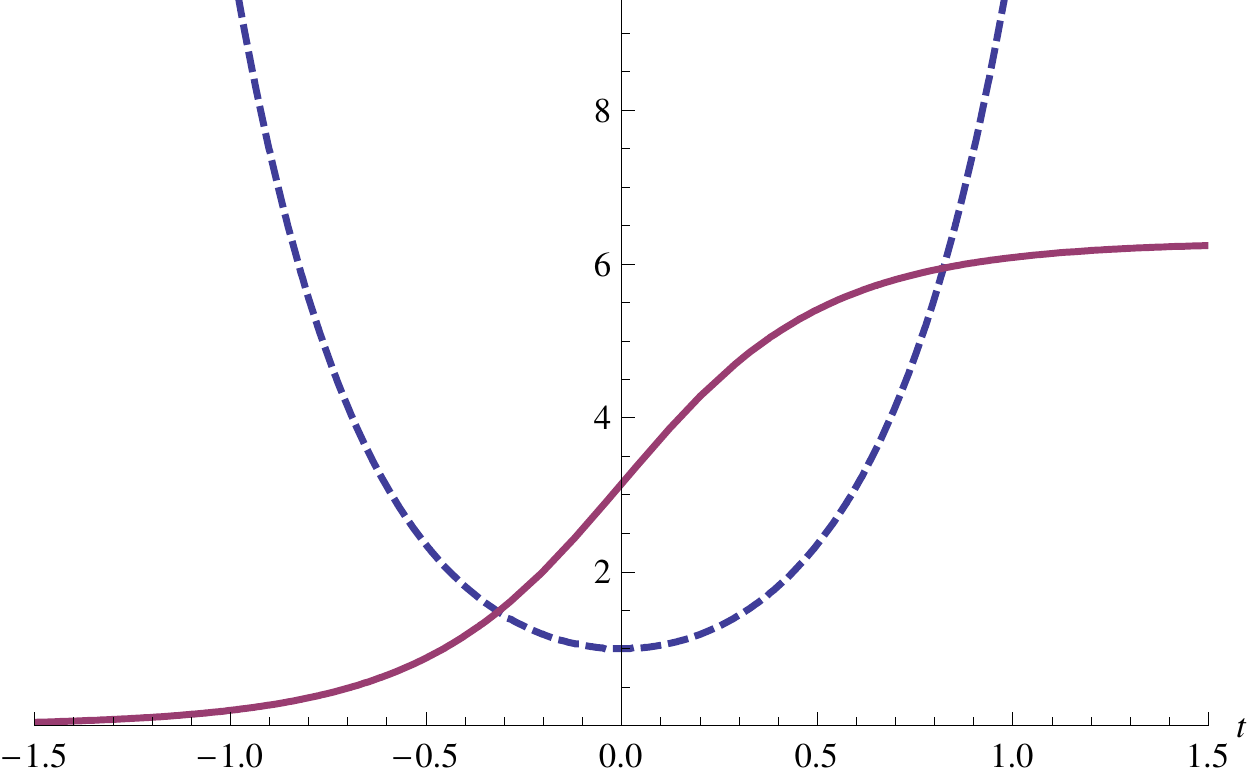}{8}{Plot of the torsion domain wall solution vs. $t$. The blue dashed line is $e^{A(t)}$ while the red solid line is $\Theta(t)$. To make the plot, we have chosen $\Theta_0=0$, $t_0=0$ and $\alpha=1$.}

Note the curvature and torsion of this solution:
\beqn
{R^\alpha}_\beta &=& -\dot F\dot A\ {\epsilon^\alpha}_{\beta\gamma}dt\w e^\gamma- a^2 e^\alpha\w e_\beta\,,\\
{R^\alpha}_0 &=& \left(\dot h+h^2\right)dt\w e^\alpha-\frac{1}{2}\dot F\dot A\ {\epsilon^\alpha}_{\beta\gamma}e^\beta\w e^\gamma\,,\\
T^\alpha&=&- \frac{1}{2}\dot F\ {\epsilon^\alpha}_{\beta\gamma}e^\beta\w e^\gamma\,,\\
T^0&=&0\,.
\eeqn
These are non-singular for all $t\in(-\infty,\infty)$. The torsion DW solution has divergent action, but this divergence is cancelled by boundary counterterms, the same counterterms which render the action of $AdS_4$ finite. To see this, the energy of the torsion DW can be computed by evaluating the Euclidean action on the solution. Introducing a cutoff at $t=\pm L$, we find
\beqn
I_{tv,on-shell}&=&4a^2\int\epsilon_{\alpha\beta\gamma}dx^\alpha\w dx^\beta\w dx^\gamma\ \int dt e^{3A(t)}\\
&=&(6\int \widehat{Vol}_3)\cdot\left(\frac{4}{3} a\alpha^3 e^{3aL}+\ldots\right)\,,
\eeqn
where the ellipsis contains terms that vanish when the cutoff is removed. As in pure $AdS_4$, an appropriate counterterm is of the form \cite{Balasubramanian:1999re,Kraus:1999di}
\beq
I_{c.t.}=-\frac{4a}{3}\int_{\pa M} \epsilon_{\alpha\beta\gamma}\tilde e^\alpha\w \tilde e^\beta\w \tilde e^\gamma\,.
\eeq
In the present case, we have such a counterterm on {\it each} asymptotic boundary, and thus we find
\beq
I_{c.t.}=-2\frac{2a}{3}\alpha^3 e^{3aL}\cdot(6\int \widehat{Vol}_3)\,,
\eeq
which exactly cancels the divergent energy of the torsion DW.

Furthermore, we note that in the Kalb-Ramond representation, the solution has
\beq
H= \dot \Theta Vol_3 =\pm 6a\alpha^3\widehat{Vol}_3\equiv \hat H \widehat{Vol}_3\,,
\eeq
where $\widehat{Vol}_3=\frac16\epsilon_{\alpha\beta\gamma}dx^\alpha\w dx^\beta\w dx^\gamma$. This corresponds to a `topological quantum number' of the kink
\beq
\int *_4H=\pm\Delta\Theta=\pm 2\pi.
\eeq

\section{The torsion domain wall as the gravity dual of parity symmetry breaking}


The holographic interpretation of the torsion DW is rather interesting.
To study this, we set to zero without loss of generality the integration constant $\Theta_0=0$ and pick the plus sign in (\ref{Pf}), (\ref{Theta}). Next we need the asymptotic expansion of the vierbein which reads
\beq
\label{FGe}
\tilde{e}^\alpha =2^{-1/3}\alpha e^{\pm a(t-t_0)}\left(1+\frac{1}{3}e^{\mp 6a(t-t_0)}+\cdots\right)dx^\alpha\,\,\, {\rm for}\,\,\, t\rightarrow \pm\infty\,.
\eeq
This shows that our solution is asymptotically anti-de Sitter for both $t\rightarrow\pm\infty$. The two asymptotic AdS spaces have the same cosmological constant. From this expansion we could read the expectation value of the renormalized boundary energy momentum tensor which would be given by the coefficient of the $e^{\pm 3at}$ term  (see e.g. \cite{MPT1,MPT2}). Such a term is missing in (\ref{FGe}), hence the expectation value of the boundary energy momentum tensor is zero. 

It is not immediately apparent how to interpret these two asymptotic regimes. Are they truly distinct, or should they be identified in some way? We note that the pseudoscalar behaves in these asymptotic regimes as
\beqn
\label{Theta+t}
\Theta(t) &\rightarrow &4e^{-3a(t-t_0)}-\frac{4}{3}e^{-9a(t-t_0)}+\cdots \,\,\, {\rm for} \,\,\, t\rightarrow -\infty \,,\\
\label{Theta-t}
\Theta(t) &\rightarrow &  2\pi -4e^{3a(t-t_0)}+\frac{4}{3}e^{9a(t-t_0)}+\cdots\,\,\,{\rm for}\,\,\ t\rightarrow +\infty\,.
\eeqn
From the above we confirm that $\Theta(t)$ is dual to a dimension $\Delta=3$ boundary pseudoscalar that we denote ${\cal O}_3$. In each one of the asymptotically AdS regimes, the leading constant behavior of $\Theta(t)$ corresponds to the source (i.e., coupling constant) for ${\cal O}_3$ and the subleading term proportional to $e^{\mp 3a(t-t_0)}$ to the expectation value $\langle{\cal O}_3\rangle$. 
The two asymptotic regimes are distinguished by the behavior of $\Theta$. In fact, the essential difference is {\it parity}.

We can now describe the holography of our torsion DW. In the $t\rightarrow -\infty$ boundary sits a three-dimensional CFT at a parity breaking vacuum state. The order parameter is the expectation value of the pseudoscalar which is $\langle {\cal O}_3\rangle =4$ in units of the AdS radius. The expectation value breaks of course the conformal invariance of the boundary theory. Then, the theory is deformed by the same pseudoscalar operator $g{\cal O}_3$ where $g$ is a marginal coupling. The torsion DW provides the holographic description of that deformation. Nevertheless, our solution should not be interpreted in terms of the usual holographic renormalization group flow. In our case, at $t\rightarrow  +\infty$ the space becomes AdS with the {\it same} radius as at $t\to-\infty$. Hence, the two boundary theories have the  same `central charges'.\footnote{We use ``central charge" in $d=3$ for a quantity that counts the massless degrees of freedom at the fixed point. Such a quantity may be taken to be the coefficient in the two-point function of the energy momentum tensor or the coefficient of the free energy density. There is no conformal anomaly in $d=3$.} 

We suggest that instead of interpreting the solution in terms of an RG flow, we should think of it as a domain wall transition between two inequivalent vacua of a single theory. This statement is supported by the behavior of $\Theta(t)$ in the two asymptotic regimes.
For  $t\rightarrow \infty$ the pseudoscalar asymptotes to the configuration (\ref{Theta-t}). The interpretation is now that when the marginal coupling takes the fixed value $g_*=2\pi$ we are back to the {\it same} CFT (i.e. having the same central charge) however in a distinct parity breaking vacuum such that $\langle{\cal O}_3\rangle =-4$. In others words, the two asymptotic AdS regimes seem to describe two distinct parity breaking vacua of the same theory. The two vacua are distinguished by the expectation value of the parity breaking order parameter being $\langle {\cal O}_3\rangle =\pm 4$. Quite remarkably, we also seem to find that starting in one of the two vacua, we can reach the other by a marginal deformation with a {\it fixed} value of the deformation parameter. 

Since the marginal operator is of dimension $\Delta=3$ and parity odd, we tentatively identify it with a Chern-Simons operator of a boundary gauge field. In this case the torsion DW induces the T-transformation in the boundary CFT \cite{Witten,LPsl2z}. In  Appendix \ref{app:C} we will argue that the 3d Gross-Neveu model coupled to abelian gauge fields exhibits a large-$N$ vacuum structure that matches our holographic findings. Although our bulk model is extremely simple to provide details for its possible  holographic dual, we regard this remarkable similarity as strong qualitative evidence that our torsion DW is the gravity dual of the `tunneling' between different parity breaking vacua in three dimensions. However, in a three-dimensional quantum field theory, we do not expect that tunneling can occur because of large volume effects, and distinct vacua remain orthogonal. Thus, referring to the torsion DW as a tunneling event should be taken figuratively. We leave to future work a more careful study of the boundary interpretation of the torsion DW solution. An interpretation will depend on the precise topology of the boundary.\cite{LNP2} Moreover, embedding our model into M-theory could provide additional clues regarding its holography.

\section{Physics in the bulk: the superconductor analogy}

The bulk interpretation of the exact solution is also interesting. Because the pseudoscalar field undergoes $\Theta(t)\to \Theta(t)+2\pi$ under $t$ goes from $-\infty$ to $+\infty$, the exact
solution corresponds to a topological kink. 
It satisfies 
\[ \int dt\dot \Theta=2\pi \]
In Figure 2, we plot the solution.
\myfig{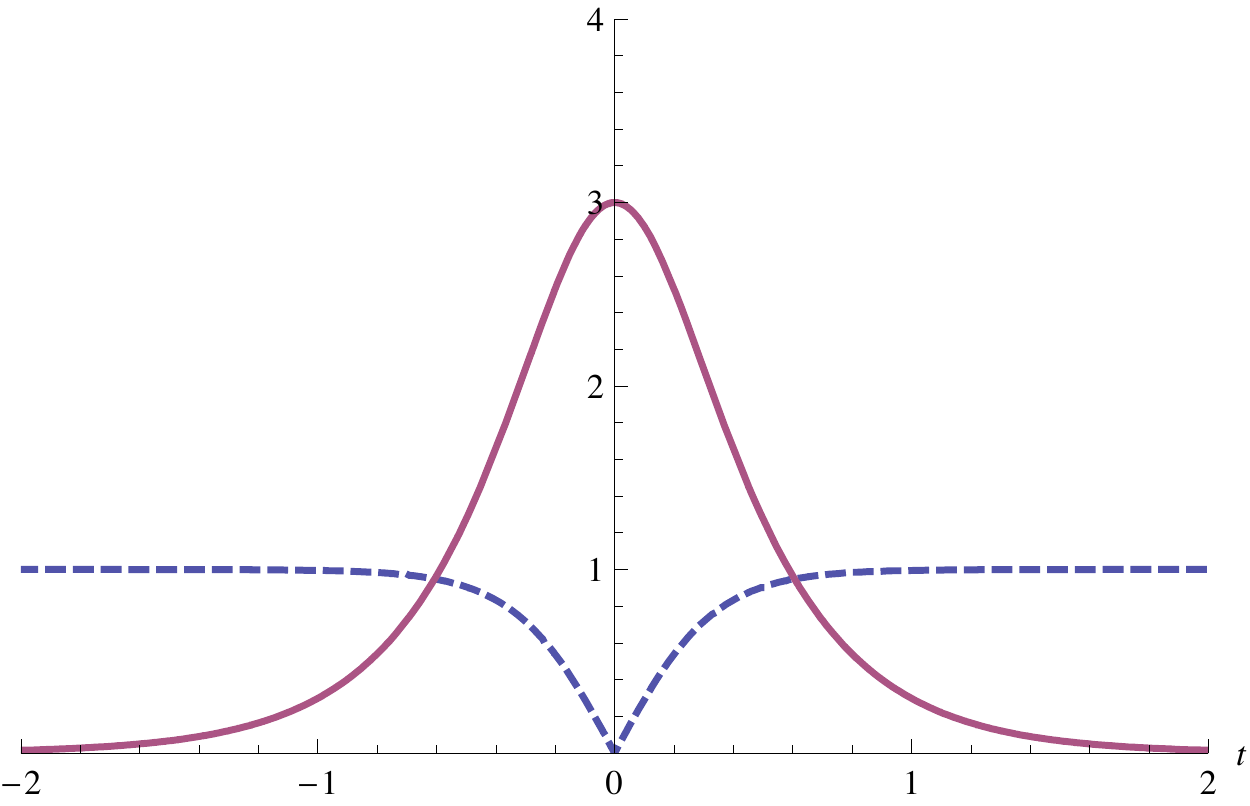}{10}{The blue dashed line is $|h(t)|$, resembling the order parameter of a superconductor, while the solid red line is $\Pi_F$, analogous to the magnetic induction of an Abrikosov DW.}

\subsection{Torsion domain wall vs Abrikosov vortex}

The gravity DW solution (\ref{eq:hdot}--\ref{Theta}) bears some resemblance to the Abrikosov vortex of superconducting systems. To avoid confusion we emphasise here that this is {\it not} a holographic interpretation i.e. the Abrikosov vortex (or more precisely, domain wall) is in the bulk. In this section, we will explore this and point out some possibly interesting features. The first thing to notice is that the plot in Fig. 2 is identical to the profile of an Abrikosov vortex (see for example Figure 5.1 in Ref. \cite{tinkham}.) The codimension differs and this is expected; the torsion DW supports a 3-form field strength in contrast to a 2-form field strength supported by the Abrikosov vortex. Nevertheless, there is a correspondence between our radial $t$-direction and the radial direction in the Abrikosov vortex, and $|h|$ and $\Pi_F$ correspond to the condensate and magnetic induction of the superconductor, respectively.
Table 1 summarizes the correspondence. 
\begin{table}[ht]
\centering
\begin{tabular}{c  c}
  \hline\hline\\
  $\h{30pt}$ \text{Abrikosov vortex}$\h{30pt}$ & $\h{30pt}$\text{Torsion DW}$\h{30pt}$ \\[1ex]
  \hline\\
  order parameter $\Phi$ & order parameter $|h|=|{\dot A}|$\\[1ex]
    $T-T_c$  & $\Lambda$  \\[1ex]
  magnetic induction B & $\Pi_F$\\[1ex]
  magnetic field H & $\hat H$\\[1ex]
  $\ZZ$-quantized magnetic flux & $\ZZ_2$-quantized
  electric flux\\[1ex]
  \hline
\end{tabular}\label{table 1}
\caption{Abrikosov vortex v.s. Torsion DW}
\end{table}
In this correspondence, since the order parameter is $h=\dot A$, the superconducting phase (constant order parameter) corresponds to $AdS_4$, while the normal phase corresponds to flat space ($h=0$). 
Far away from the core of the torsion DW, the geometry is asymptotically $AdS$, but at the core the spatial slice (at $t\to t_0$) becomes flat. To see this, note that if we think of the system as a pseudoscalar coupled to torsionless gravity, the torsion DW has ${\cc{\omega}^\alpha}_\beta=0$ and ${\cc{\omega}^\alpha}_0= \dot A\tilde e^\alpha$, and so
\beqn
{\cc{R}^\alpha}_\beta &=& - h^2 \tilde e^\alpha\w\tilde e_\beta\,,\\
{\cc{R}^\alpha}_0&=&(\dot h+h^2) dt\w\tilde e^\alpha\,,\\
\cc{T}^\alpha &=&0\,.
\eeqn
Thus, at the core, we find that the Riemann tensor has components
\begin{align}
  {R^{\alpha}}_{0\alpha 0} &\to -3 a^2 \alpha\,,\\
  {R^\alpha}_{\beta\alpha\beta} &\to 0\,.
\end{align}
This behavior is in line with an Abrikosov vortex in which there is normal phase at the core and superconducting phase away from the core. 

The analogue of the magnetic field is what we have called $\hat H$, proportional to the constant $\alpha^3$. In the DW, the magnetic induction, analogous to $\Pi_F$, has a penetration length $\lambda\sim 1/3a$, and the coherence length of the order parameter is $\xi\sim 1/6a$. The penetration and coherence lengths are obtained by looking at the exponential fall-off of these quantities in the core of the DW, away from their values in the superconducting phase.

The torsion DW also has a quantized flux $\int *_4 H=\Delta\Theta=2\pi$.  This flux is independent of any parameters of the solution and of any rescaling of fields in the theory. Thus, this is an analogue of the quantized magnetic flux in superconductivity.

Finally, note the following interesting feature. If we take a derivative of the second equation in (\ref{eq:hdot}), we arrive at
\begin{equation}\label{dot-h-eq}
\ddot{h} - 6\Lambda h - 18 h^3 = 0\,.
\end{equation}
This looks like a Landau-Ginzburg equation of motion of an effective $\phi^4$ theory. This leads us to interpret 
\beq
\Lambda =-3\sigma_\perp a^2=-3\sigma_\perp\frac{1}{L^2}\sim T_c-T
\eeq
 Of course, there is no real temperature in the case of the torsion DW, but we note that this implies that the penetration and coherence lengths diverge as 
 \beq
 (T-T_c)^{1/2}
 \eeq
 i.e.  with a mean field theory critical exponent $1/2$, as in superconductivity.

\subsection{Domain wall condensation: a prelude to Kalb-Ramond superconductivity?}

In the last section, we noted that there is a strong analogue between the torsion DW solution and superconductivity. It is intriguing to carry the analogy further and consider multi-DW configurations. We have noted that at the core of the torsion DW, the spatial sections are flat. Thus, one might imagine that if it was favourable for torsion DWs to condense, as DWs do in Type I superconductors, then finite regions of normal phase (corresponding to $\Lambda=0$) would be obtained. We will argue below that this can in fact occur, although the system appears not to be unstable. 

To understand the physics involved, the first step is to consider a configuration of two DWs. In the superconductivity literature, this is a standard computation. One takes two DWs separated by a distance $\ell$ and computes the Euclidean action. More precisely, we will treat this here as follows. Put the system in a box by restricting the $t\in[-L,L]$. In such a case, a DW located at $t_0$ has on-shell action
\beq
e^{A(t-t_0)}=\alpha(2\cosh3a(t-t_0))^{\frac{1}{3}}\Rightarrow I^L_{on-shell} =\left(6\hat{Vol}_3\right)\frac{4}{3}a\a^3e^{3a(L-t_0)}\,.
\eeq
We then consider a piecewise solution of two DWs located at $t=\pm\frac{L}{2}$ This is not a solution of the e.o.m. because solutions of non-linear equations cannot be simply superimposed i.e. it fails at the midpoint between the DWs. However, if we simply evaluate the Euclidean action, we find 
\beq
I^{\pm(L/2)}_{on-shell}=\left(\frac{4}{3}a^2\hat{Vol}_3\right)4a\a^3\sinh\frac{3aL}{2}\,.
\eeq
Note that this is positive, so one might naively conclude that the DWs repel each other. However, recall that the DW profile exists not in flat space-time, but in the metric given by (\ref{eq:expA}), which rises asymptotically. As a result, as we move the DWs further apart, there is a corresponding rise in the metric between the DWs. So, we should directly evaluate the force at a point $t=\ell$ as
\beq
F=-\frac{dI^{\pm(L/2)}}{d \ell}\propto \cosh(3a\ell/2) <0\,.
\eeq 
Thus we conclude that the DWs in fact {\it attract} each other. In the superconducting analogue we would conclude that we have a {\it Type I superconductor} where the vortices clump together forming (potentially) finite regions of normal phase. In such a superconductor, the number of vortices is determined by the total magnetic flux.

We now describe the analogous situation in our gravitational system. We have noted that the constant $\hat H$ plays the role of the external magnetic induction, while $H$ is the magnetic field, varying within the DW, with $\Delta\Theta=\int *_4 H$. Following the superconducting analogue, if we put the system in a box of size $2L$ (that is we impose a cutoff on each AdS asymptotic) the flux conservation equation is of the form
\beq\label{fluxquant}
\Delta\Theta = 2L\hat H
\eeq
The DWs carry the flux in the superconductor, and so it is natural to ask what is the lowest energy configuration satisfying (\ref{fluxquant})? To analyze this, consider an array of $n$ DWs in a region of size $L_0$. We take the DWs to be equally spaced, as one can show that deviating from such a configuration causes a rise in energy. For such a configuration, the flux quantization condition (\ref{fluxquant}) gives a relation between $n$,$L_0$ and $\hat H$. Such a representative curve is shown in Fig. 3 
\myfig{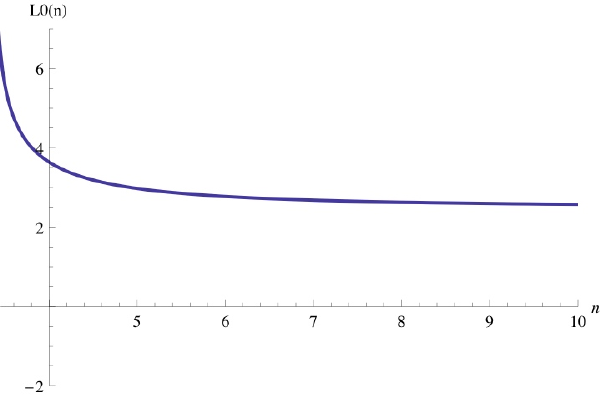}{9}
{Size of normal state droplet vs. $n$ for multi-domain walls.}

If we solve this equation for $L_0$ as a function of $n$ and $\hat H$, we can then compute the energy as a function of $n$. One obtains a curve as in Fig. 4. 

\myfig{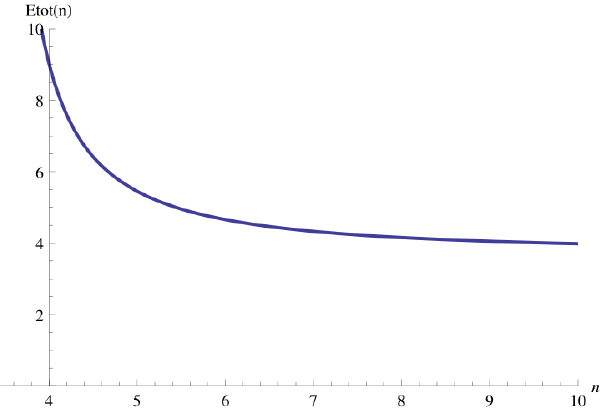}{9}{Energy vs. $n$ for multi-domain walls.}

One notes that the energy is minimized for large $n$, and in that case, the size $L_0$ asymptotes to a fixed value, which is found to be
\beq
L_0=\frac{\hat H }{6a}\cdot 2L = \alpha^3\cdot 2L 
\eeq
We conclude that the preferred configuration, given a fixed external flux, is a continuum of DWs arrayed over a finite size region. Within this droplet, the system is in the normal phase. We have noted that the DW core is spatially flat, and so we surmise that within the droplet, the space-time is flat. The asymptotic value of energy in Figure \ref{fig:mvEtot.pdf} is precisely minus that contributed by the cosmological constant. Again, the size of the droplet is set by the value of the external $H$-flux, and the boundary conditions are AdS. Note that for a fixed cutoff, there is a critical field (given by $\hat H=6a$) for which the entire spacetime is flat.

The result above motivate us to take a further step and suggest  rather appealing physical picture that we may call {\it Kalb-Ramond superconductivity}. Here we present a qualitative description of Kalb-Ramond superconductivity. A detailed description will appear shortly in \cite{LNP2}. 
As discussed above, {\it minus} the cosmological constant may be interpreted as $T-T_c$. This, together with the interpretation of $H$ as a magnetic induction and $\hat{H}=6a$ as a critical magnetic field above which the spacetime is flat due to DW condensation, motivates to draw a $-\Lambda , *H$ graph analogously with the $T-T_c, ${\it magnetic field} graph in superconductivity. To do so, we need to consider all known solutions of an Einstein-axion system in $d=4$ with and without cosmological constant. For $\Lambda=0$ we recall the axionic DW solutions of \cite{GS}. For $\Lambda<0$ DW wormhole-like solutions were found in \cite{Gutperle}.

These solutions have the following properties (to be discussed in more detail in \cite{LNP2}):

The $\Lambda=0$ wormhole solution \cite{GS} is asymptotically flat. Its magnetic flux is proportional to an arbitrary constant $g$ which was speculated to be quantized via a string theory embedding of the Einstein-axion system. in \cite{GS}. Its electric flux is $\ZZ_2$-quantized to  $\pm 3\pi$, but the electric field varies along the wormhole.  

The $\Lambda<0$ wormhole solution \cite{Gutperle} is asymptotically AdS. Its magnetic flux is also proportional to the arbitrary constant $g$, and hence possibly quantized in a string theory embedding. Quite intriguingly, the electric flux on the wormhole solution interpolates between the value $\pm 3\pi$ for $g\rightarrow 0$ and $\pm 2\pi$ as $g\rightarrow \infty$. Moreover,  there is a lower bound for magnetic field of the wormhole solution. Remarkably, this lower bound is the value $6a$ and coincides with the maximal magnetic field of the torsion DW. 

In Fig. 5 we sketch the possible phase diagram of the Kalb-ramond superconductor. Our torsion DW solution seems to play the role of the superconducting phase, while the wormhole solutions of \cite{GS} and \cite{Gutperle} appear to correspond to the normal phase.

\myfig{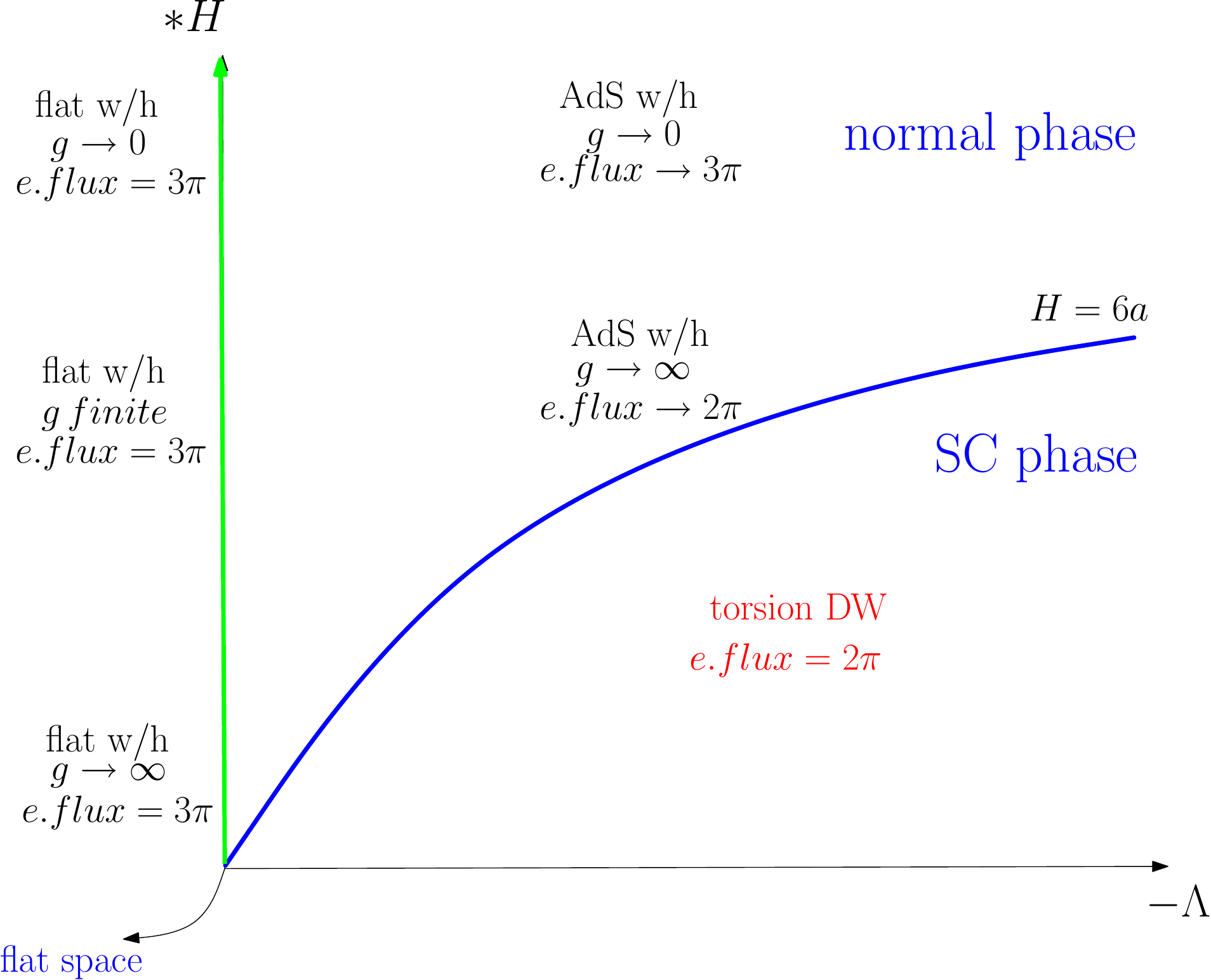}{12}{A sketch of the phase diagram for a Kalb-Ramond superconductor.}

\section{Conclusions}

In this work we have presented in detail a simple toy model, the Nieh-Yan model, where torsion enters through the spacetime dependence of the coupling constant of the Nieh-Yan topological invariant. Although we have discussed the model directly in terms of torsion, it can classically be put into equivalent forms as either a massless pseudoscalar or a Kalb-Ramond field coupled to gravity. 
The model has an interesting and non-trivial holographic interpretation. 
In particular, we have shown that it possesses an exact bulk solution in Euclidean signature, termed the torsion DW, having two asymptotically AdS$_4$ regimes, while the pseudoscalar acquires a kink profile. We have argued then that the holographic interpretation of this torsion DW is a three-dimensional CFT with two distinct parity breaking vacua. Moreover, our bulk solution may imply that the deformation by a classically marginal pseudoscalar with a fixed  coupling constant induces a transition between the two parity breaking vacua separated by a domain wall, which would be at infinity in the boundary components.\cite{LNP2} Remarkably, this qualitative behaviour is seen already in the three-dimensional Gross-Neveu model coupled to $U(1)$ gauge fields. The economy of our bulk model does not allow a detailed identification of the bulk and boundary theories, nevertheless  we believe that our results provide a strong base where an exact bulk/boundary dictionary for AdS$_4$/CFT$_3$ can be based. A further rather intriguing property of the torsion DW is that it can be mapped into the standard Abrikosov vortex of superconductivity. Such a map identifies flat spacetime with a superconductor's normal phase, while AdS is identified with a superconducting phase. The cosmological constant would then measure the deviation from the `critical temperature'. A phenomenon of DW condensation is found, similar to the analogous case in type I superconductors.
Finally, we have briefly discussed  a picture of "Kalb-Ramond superconductivity" that emerges if we view in a unified way all known 4-dimensional Euclidean solutions of the Einstein-axion system. This picture will be furhter analysed in \cite{LNP2}. 

Our results indicate that the torsional degrees of freedom of four dimensional gravity can provide holographic descriptions for a number of interesting properties of 3d critical systems. It would be interesting to extend our analysis to more elaborate models where more torsional degrees of freedom become dynamical. It is also of interest to discuss whether our simple model can be embedded into M-theory.

\subsection*{Acknowledgments}
I thank the organizers and in particular Lefteris Papantonopoulos for the organization of a top quality workshop and the invitation to present this talk. This work is partially supported by the FP7-REGPOT-2008-1 "CreteHEPCosmo" No 228644 and also by the University of Crete ELKE grant with KA 2745. I wish to thank R. G. Leigh and N. N. Hoang for the very fruitful and pleasant collaboration that has led to the results presented in this report.

\appendix
\renewcommand{\theequation}{\thesection.\arabic{equation}}
\renewcommand{\thesubsection}{\thesection.\arabic{subsection}}
\setcounter{equation}{0}
\setcounter{subsection}{0}


%


\section{Parity breaking in three dimensions}\label{app:C}

Consider the 3d Gross-Neveu model coupled to abelian gauge fields. The Euclidean action is\footnote{We use $\bar{\psi}^i$, $\psi^i$ ($a=1,2,...,N$) two-component Dirac fermions. The $\gamma$-matrices are defined in terms of the usual Pauli matrices as $\gamma^i=\sigma^i$ $i=1,2,3$.}
\beq
\label{GNaction}
I=-\int d^3x\left[\bar{\psi}^a\left(\slash\!\!\!\partial-{\rm i}e\slash \!\!\!\!A\right)\psi^a +\frac{G}{2N}\left(\bar{\psi}^a\psi^a\right)^2+\frac{1}{4M}F_{\mu\nu}F_{\mu\nu}\right]\,.
 \eeq
$M$ is an UV mass scale.   Introducing the usual Lagrange multiplier field $\sigma$, whose equation of motion is $\sigma =\frac{-2G}{N}\bar{\psi}^a\psi^a$ we can make the action quadratic in the fermions
 \beq
 \label{GNactionsigma}
 I=-\int d^3x\left[\bar{\psi}^a\left(\slash\!\!\!\partial+\sigma-{\rm i}e\slash \!\!\!\!A\right)\psi^a -\frac{N}{2G}\sigma^2-\frac{1}{4M}F^{\mu\nu}F_{\mu\nu}\right]\,.
 \eeq
The model possesses two parity breaking vacua distinguished by the value of the pseudoscalar order parameter $\langle\sigma\rangle$. This is seen as follows: switching off the gauge fields momentarily one integrates over the fermions to produce a large-$N$ effective action as
\beq
\label{GNeffact}
{\cal Z} = \int ({\cal D}\sigma)e^{N\left[\Tr\log\left(\slash\!\!\!\partial +\sigma\right)-\frac{1}{2G}\int d^3x\sigma^2\right]}
\,.
\eeq
The path integral has a non-zero large-$N$ extremum $\sigma_*$ found by setting 
$\sigma =\sigma_* +\frac{1}{\sqrt{N}}\lambda$ 
\beq
\label{GNeffactN}
{\cal Z} =\int ({\cal D}\lambda) e^{N\left[\Tr\log\left(\slash\!\!\!\partial +\sigma_*\right)-\frac{1}{2G}\int d^3x \sigma_* +\frac{1}{\sqrt{N}}\left\{ \Tr\frac{\lambda}{\slash\!\!\!\partial+\sigma_*}-\frac{\sigma_*}{G}\int d^3x\lambda\right\}+O(1/N)\right]}
\eeq
The term in the curly brackets is the gap equation. To study it one considers a uniform momentum cutoff $\Lambda$ to obtain
\beq
\label{GNgap}
\frac{1}{G}=\int^\Lambda\frac{d^3p}{(2\pi)^3}\frac{2}{p^2+\sigma_*^2}=(\Tr 1)\left[\frac{\Lambda}{\pi^2}-\frac{|\sigma_*|}{\pi^2}\arctan\frac{\Lambda}{|\sigma_*|}\right]\,.
\eeq
Defining the critical coupling as
\beq
\label{GNcritcoupl}
\frac{1}{G_*}=\frac{\Lambda}{\pi^2}\,,
\eeq
(\ref{GNgap}) possesses a non-zero solution for $\sigma_*$ when $G>G_*$ given by
\beq
\label{sigma*}
|\sigma_*|=\frac{2\pi}{G}\left(\frac{G}{G_*}-1\right)\equiv m\,.
\eeq
The two distinct parity breaking vacua then have
\beq
\label{u}
\sigma_*=-\frac{2G}{N}\langle\bar{\psi}^a\psi^a\rangle =\pm m
\,.
\eeq
Going back to (\ref{GNactionsigma}) one can tune $G>G_*$ and start in any of the two parity breaking vacua. Suppose we start from $\sigma_*=+m$. To leading order in $N$ we have 
\beqn
\label{GNsigmam}
{\cal Z}&=&\int ({\cal D}A_\mu)({\cal D}\bar{\psi}^a)({\cal D}\psi^a)\exp[{\cal S}]\\
{\cal S}&=&\int d^3x\left[\bar{\psi}^a\left(\slash\!\!\!\partial +m-{\rm i}e\slash\!\!\!\! A\right)\psi^a-\frac{N}{2G}m^2+O(1/\sqrt{N})-\frac{1}{4M}F^{\mu\nu}F_{\mu\nu}\right]\nonumber 
\eeqn
As is well known \cite{Semenoff,Redlich} for $N$ fermions the path integral (\ref{GNsigmam}) yields an effective action for the gauge fields which for low momenta is dominated by the Chern-Simons term i.e.
\beq
\label{GNCS}
{\cal Z} \approx \int e^{S_{CS}}
\,,
\eeq
with
\beq
\label{CS}
S_{CS} ={\rm i}\frac{ke^2}{4\pi}\int d^3x \epsilon^{\mu\nu\rho}A_\mu\partial_\nu A_\rho\,,\,\,\,\,\,\,\,\, k=\frac{N}{2}
\,.
\eeq
Had we started from the $\sigma_*=-m$ vacuum, we would have found again (\ref{GNCS})-(\ref{CS}), however with $k=-\frac{N}{2}$, i.e. the vacuum with $\sigma_*=-m$ yields an effective Chern-Simons action with $k=-\frac{N}{2}$.

 Consider now {\it deforming} the action (\ref{GNsigmam}) by the Chern-Simons term with a fixed coefficient as
 \beqn
 \label{GNsigmamDef}
 {\cal Z}_{deform}&=&\int ({\cal D}A_\mu)({\cal D}\bar{\psi}^i)({\cal D}\psi^i)\exp[{\cal S}_{def}]\\
 {\cal S}_{def}&=&{\cal S}-{\rm i}q\int d^3x\epsilon^{\mu\nu\rho}A_\mu\partial_\nu A_\rho \nonumber 
\eeqn
If $q$ is fixed to
\beq
\label{q}
q=\frac{Ne^2}{4\pi}\,,
\eeq
the effective action for the gauge fields resulting from the fermionic path integrals in (\ref{GNsigmamDef}) is going to be {\it equal}  to the one obtained had we started at the  $\sigma_*=-m$ vacuum. In other words, deforming the $\sigma_*=+m$ vacuum with a Chern-Simons term with a fixed coefficient is equivalent to being in the $\sigma_*=-m$ vacuum. This is exactly analogous to the holographic interpretation of our torsion DW.

%




%
\providecommand{\href}[2]{#2}\begingroup\raggedright\endgroup

\end{document}